\documentclass[preprint,eqsecnum,aps,showpacs]{revtex4}

\begin{document}


\newcommand{\xbf}[1]{\mbox{\boldmath $ #1 $}}

\newcommand{\sixj}[6]{\mbox{$\left\{ \begin{array}{ccc} {#1} & {#2} &
{#3} \\ {#4} & {#5} & {#6} \end{array} \right\}$}}

\newcommand{\threej}[6]{\mbox{$\left( \begin{array}{ccc} {#1} & {#2} &
{#3} \\ {#4} & {#5} & {#6} \end{array} \right)$}}

\title{Baryon Charge Radii and Quadrupole Moments in the $1/N_c$
Expansion: The 3-Flavor Case}

\author{Alfons J. Buchmann}
\email{alfons.buchmann@uni-tuebingen.de}

\affiliation{Institute for Theoretical Physics, University of
T\"{u}bingen, D-72076, Germany}

\author{Richard F. Lebed}
\email{Richard.Lebed@asu.edu}

\affiliation{Department of Physics and Astronomy, Arizona State University,
Tempe, AZ 85287-1504}

\date{July, 2002}

\begin{abstract}
We develop a straightforward method to compute charge radii and
quadrupole moments for baryons both with and without strangeness, when
the number of QCD color charges is $N_c$.  The minimal assumption of
the single-photon exchange ansatz implies that only two operators are
required to describe these baryon observables.  Our results are
presented so that SU(3) flavor and isospin symmetry breaking can be
introduced according to any desired specification, although we also
present results obtained from two patterns suggested by the quark
model with gluon exchange interactions.  The method also permits to
extract a number of model-independent relations; a sample is
$r^2_\Lambda / r_n^2 = 3/(N_c+3)$, independent of SU(3) symmetry
breaking.
\end{abstract}


\maketitle
\thispagestyle{empty}

\newpage
\setcounter{page}{1}

\section{Introduction}\label{sec:intro}

Between the static properties of hadrons, {\it e.g.}, masses,
electromagnetic moments, matter and charge radii, and low-energy
dynamical properties such as scattering lengths and decay rates, a
wide variety of accurately measured data lies outside the perturbative
range of QCD.  While techniques such as lattice gauge theory, QCD sum
rules, and wide variety of models have been developed to study this
extremely interesting energy regime, the only known perturbative
approach for strongly-interacting Yang-Mills theories that holds at
all energies is the $1/N_c$ expansion~\cite{tHooft}, where $N_c$ is
the number of color charges.

The baryon sector is particularly amenable to this expansion, since
baryons with $N_c$ colors contain $N_c$ valence quarks.  Then, each
additional quark participating in an interaction (specified by an
operator with known transformation properties under spin and flavor)
brings in a factor of the QCD coupling constant $\alpha_s \propto
1/N_c$ due to the requirement of one or more gluons to connect this
quark to the interaction.  An operator in which $n$ quarks interact is
called an $n$-body operator; its suppression in powers of $1/N_c$
tends to increase as $n$ increases.  This ``operator method'' has
recently been used to study a wide variety of baryon
observables (see Ref.~\cite{BHL} for a recent list of references).
The strength of the $1/N_c$ expansion is that additional gluons do not
spoil this counting, and the only other powers of $N_c$ that need be
taken into account arise from the combinatorics of the quarks.  It
should also be noted that the glue and sea quark pairs in the baryon,
not just the valence quarks, are subsumed by the operator
method~\cite{BL1}.

This work extends the results of two of our recent papers on baryon
observables in the $1/N_c$ expansion, regarding charge
radii~\cite{BL1} and quadrupole moments~\cite{BHL}, to baryons with
nonzero strangeness.  These two types of observables are studied
here in a single paper because the method of calculation is, as seen
below, very similar in the two cases.  As mentioned in the previous
works, a treatment of the strange sector was not undertaken at the
time of their writing because the group theory in the 3-flavor sector
is more involved and warrants a separate treatment.  The operator
method calculation specific to the case $N_c=3$, otherwise known as
the general QCD parameterization~\cite{Morpurgo} was carried out for
baryon quadrupole moments in~\cite{BH2}; in that case the calculation
can readily be carried out for states with or without strangeness
since it is relatively simple to perform calculations using the full
baryon spin-flavor wave functions with only 3 quarks.

Here we demonstrate that the solution of the full 3-flavor $N_c$-quark
problem can nonetheless be handled entirely using SU(2) Clebsch-Gordan
(CG) coefficients, ultimately because the strange states are related
to nonstrange states through the SU(2) $U$- and $V$-spin subgroups of
SU(3)~\cite{Car}.  In the case of the ground-state baryons, {\it
i.e.}, those belonging to the large-$N_c$ generalization of the SU(6)
{\bf 56}-plet, the total symmetry of the wave function under
simultaneous exchange of spin and flavor indices simplifies this
procedure considerably.

In this paper we focus only on those aspects of the problem unique to
the strange sector; the reader is directed to Refs.~\cite{BHL} and
\cite{BL1} for a more thorough discussion of details of the $1/N_c$
operator method, as well as outlook for the experimental measurement
of baryon quadrupole moments.  The remainder of this paper is
organized as follows: In Sec.~\ref{CR} we briefly discuss the
experimental situation regarding measurement of baryon charge radii.
Section~\ref{ops} presents the details of the 3-flavor calculation;
the most important feature of the analysis of \cite{BHL} and
\cite{BL1} survives, namely, that only one operator appears at the
2-body level and one at the 3-body level, leading to a large number of
constraints between the observables.  Section~\ref{results} presents a
sample of the plethora of results obtained from this calculation, {\it
e.g.}, by including frequently used patterns of SU(3) flavor symmetry
breaking.  Section~\ref{concl} summarizes.  The Tables and Appendices
contain the extensive results of the calculation in forms designed to
be most useful to interested researchers.

\section{Charge Radii: Prospects for Measurement} \label{CR}

The mean-square charge radius $r^2_B$ of a baryon $B$ is defined
through the elastic Sachs charge form factor $F(q^2)$, a function of
the photon 4-momentum transfer $q^2$ often denoted by $G_E (q^2)$, by
the relation
\begin{equation}
r^2_B \equiv -6 \frac 1 Q \left. \frac{dF(q^2)}{dq^2} \right|_{q^2=0}
,
\end{equation}
while the total charge $Q = F(0)$ is omitted if $Q=0$.
 
The operator giving rise to charge radii is a scalar that does not
change electric charge or strangeness.  Thus, it connects only states
with the same total $J$, $I_3$, and $Y$.  However, the operator can
connect states of different isospin $I$ since the electromagnetic
interaction does not conserve this quantum number.  Each baryon state
thus has a charge radius, and in addition there exists a
$r^2_{\Sigma^0 \Lambda}$ transition charge radius.

At the time of this writing, only the $p$, $n$, and $\Sigma^-$ charge
radii have been measured.  A charge radius is the first nontrivial
moment of a Coulomb monopole ($C0$) transition amplitude, and thus
charge radii of hadrons are typically measured through their coupling
to Coulomb photons in elastic electron scattering.

In the case of the proton, a dedicated measurement of the elastic
electron-proton cross section at very low momentum
transfer~\cite{SSBW} led to the proton root mean-square charge radius
$r_p \equiv \sqrt{r^2_p} = 0.862(12)$ fm [$r^2_p = 0.743(21)$ fm$^2$],
a value that is considerably larger than the famous dipole result of
$r_p = 0.80$ fm obtained by Hofstadter {\it et al.}~\cite{Hof}.  A
dispersion-theoretical analysis of electron scattering data available
through 1995~\cite{MMD} gives the result $r^2_p = 0.717(15)$ fm$^2$.
However, measurements using the hydrogen Lamb shift as an alternative
source of information on the proton size tend to produce larger
values.  For example, Ref.~\cite{Udem} extracts a value of $r_p =
0.890(14)$ fm [$r^2_p = 0.792(25)$ fm$^2$], while Ref.~\cite{MR}
extracts a value of $r^2_p = 0.780(25)$ fm$^2$ from the experiment of
Ref.~\cite{Schwob}.  A reanalysis~\cite{Rose} that includes Coulomb
and recoil corrections for $A>1$ targets produces a result that
accommodates both types of data:
\begin{equation} \label{rpmeas}
r^2_p = 0.779(25) \ {\rm fm}^2 ,
\end{equation}
which we use in our analysis.  The other available experimental values
are
\begin{equation} \label{rnmeas}
r^2_n = -0.113(3)(4) \ {\rm fm}^2~\mbox{\cite{rn2}} ,
\end{equation}
and the very recently published result
\begin{equation} \label{rsigmes}
r^2_{\Sigma^-} = 0.61(12)(9) \ {\rm fm}^2~\mbox{\cite{SELEX}}.
\end{equation}
The measurement of $r^2_{\Sigma^-}$ in particular suggests the
possibility of measuring the charge radii of other long-lived strange
baryons, the $\Sigma^+$, $\Lambda$, $\Xi^-$, $\Xi^0$, and $\Omega^-$.
In such cases the impediments are experimental in nature, particularly
the problem of producing a beam of baryons of sufficient quality that
electron scattering events can be separated from background events.
But nonetheless one can anticipate such difficulties being overcome in
the future.

For the spin-3/2 resonances and $\Sigma^0$, however, the short
lifetimes mean that charge radii can only be observed in off-shell
processes.  For example, the $\Delta^{++}$ charge radius could be
extracted in principle through the process $\pi^+ p \to \pi^+ p \, e^-
e^+$, but in this case one would need to use model dependence in
separating resonant from continuum $\pi^+ p$ scattering, as well as
isolate the source of the virtual photon $\gamma^* \to e^+ e^-$ as
coming solely from the $\Delta$.  Processes with real photons, which
for example can be used to measure magnetic moments, cannot be
utilized to probe the Coulomb transitions from which charge radii are
extracted.  The situation is very similar to that described in
Ref.~\cite{BHL} for the baryon quadrupole moments.  Therefore, our
results divide into two categories: Predicitions for observables that
may be measured in the near-to-medium future, for which the quality of
our results can be checked, and those that can only be predicted but
not measured any time soon.

\section{Details of the Calculation} \label{ops}

\subsection{Constructing the States}

Naively, one would expect that SU(3) CG coefficients are required to
construct a state with strangeness.  This potential complication can
be avoided by noting that the strange states are related to the
nonstrange states via $U$- and $V$-spin SU(2) subgroups of SU(3).  In
an arbitrary SU(3) representation, one method of calculating matrix
elements is to start with the nonstrange states and apply the lowering
operators $U_-$ and $V_-$, imposing along the way orthogonality in
isospin among otherwise degenerate states such as $\Sigma^0$ and
$\Lambda$, which is precisely the same approach as used to obtain
flavor eigenstates.

However, this method does not exploit the full symmetry available to
the $J^P = 1/2^+$ and $3/2^+$ ground-state baryon SU(3) multiplets (We
do not call them ``octet'' and ``decuplet,'' since the corresponding
multiplets for arbitrary $N_c$ are much larger~\cite{reviews}).  These
flavor multiplets belong to the large-$N_c$ analogue of the SU(6)
{\bf 56}-plet, which is completely symmetric under simultaneous
exchange of spin and flavor indices.  That is to say, the $I$-, $U$-,
and $V$-spin symmetries are correlated with the SU(2) spin symmetry in
a particularly convenient way, as we now explore.

Let $N_\alpha$ denote the number of valence quarks of type $\alpha$ in
the baryon; then $N_c = N_u + N_d + N_s$ and $I_3 = (N_u - N_d)/2$.
Since the spin-flavor wave function is completely symmetric, the spin
wave function of quarks of type $\alpha$ within the baryon must be
completely symmetric, which means that the $N_\alpha$ spin-1/2 quarks
of type $\alpha$ carry the maximum possible spin value, $S_\alpha =
N_\alpha/2$.  This information alone allows to analyze the nonstrange
states.  In the strange baryon case, note that each physical baryon
state is still specified by its isospin $I$, which uniquely determines
the symmetry of the flavor wave function carried by just $u$ and $d$
quarks.  Owing to the complete symmetry under spin $\times$ flavor,
the spin wave function carried by the $u$ and $d$ quarks together must
possess exactly the same symmetry properties as the corresponding
flavor wave function, and hence, the spin ${\bf S}_{ud} \equiv {\bf
S}_u + {\bf S}_d$ and isospin ${\bf I}$ have the same eigenvalue,
$S_{ud} = I$.

But now one need only combine the state of $ud$ combined spin $S_{ud}
= I$ and isospin quantum numbers $I, I_3$ with the symmetrized strange
quarks carrying spin $S_s$ to obtain the complete state with spin
eigenvalues $J, J_3$, where ${\bf J} = {\bf S}_{ud} + {\bf S}_s$.  To
be precise,
\begin{eqnarray}
\left| J J_3; \, I I_3 \; (S_u \, S_d \, S_s) \right> & = &
\sum_{S_{ud}^z , \, S_s^z} \left( \begin{array}{cc}
I & S_s \\ S_{ud}^z & S_s^z \end{array} \right| \left.
\begin{array}{c} S \\ S_z \end{array} \right)
\sum_{S_u^z , \, S_d^z} \left( \begin{array}{cc}
S_u & S_d \\ S_u^z & S_d^z \end{array} \right| \left.
\begin{array}{c} I \\ S_{ud}^z \end{array} \right) \left| S_u \, S_u^z
\right> \left| S_d \, S_s^z \right> \left| S_s \, S_s^z \right> ,
\nonumber \\ & & \label{ket}
\end{eqnarray}
where the parentheses denote CG coefficients.  Now, in order to
compute the matrix elements of any particular operator, one need only
sandwich it between a bra and ket of the form of Eq.~(\ref{ket}) and
use the Wigner-Eckart theorem.

\subsection{The Operator Basis}

In the analysis of both charge radii and quadrupole moments we use the
single-photon exchange assumption, {\it i.e.}, that the photon probing
these baryon observables couples to only one quark line within the
baryon.  Although physically rather mild, this assumption drastically
reduces the number of distinct operators that need be considered.
(However, note that the full basis of operators can certainly be used,
as shown for the charge radii~\cite{BL1} or for the magnetic and
quadrupole moments~\cite{JJM}.)  Using indices $i,j,k$ to indicate
quarks within the baryon, the most general operator expansion for the
charge radius operator out to the 3-body level---all that is necessary
since 4-,5-, etc.\ body operators acting upon physical baryons are
linearly dependent on those at lower order---reads
\begin{equation} \label{chradop}
\left. -6 \frac{dF(q^2)}{dq^2} \right|_{q^2=0} = A \, \sum_i^{N_c} a_i
+ \frac{B}{N_c} \sum_{i \neq j}^{N_c} b_i \, c_j \; \xbf{\sigma}_i \!
\cdot \! \xbf{\sigma}_j + \frac{C}{N_c^2} \! \sum_{i \neq j \neq k}
d_i \, e_j \, f_k \; \xbf{\sigma}_i \! \cdot \! \xbf{\sigma}_j ,
\end{equation}
where the coefficients $a,b,c,d,e,f$ depend only on quark flavor; to
obtain the results of Ref.~\cite{BL1}, set $a_i = b_i = f_i = Q_i$,
where $Q_i$ is the charge of the $i$th quark, and $c_i = d_i = e_i =
1$.  The corresponding expansion for the quadrupole moment operator
reads
\begin{equation} \label{quadop}
{\cal Q} = \frac{B^\prime}{N_c} \sum_{i \neq j} b^\prime_i \,
c^\prime_j \, ( 3 \sigma_{iz} \sigma_{jz} - \xbf{\sigma}_i \! \cdot \!
\xbf{\sigma}_j ) + \frac{C^\prime}{N_c^2} \! \sum_{i \neq j \neq k}
d^\prime_i \, e^\prime_j \, f^\prime_k \; \xbf{\sigma}_i \! \cdot \!
\xbf{\sigma}_j ,
\end{equation}
where the coefficients $b^\prime, c^\prime, d^\prime, e^\prime,
f^\prime$ again depend only on quark flavor, and the results of
Ref.~\cite{BHL} may be obtained by setting $b^\prime_i = f^\prime_i =
Q_i$, $c^\prime_i = d^\prime_i = e^\prime_i = 1$.  Note that most
models use a universal form of SU(3) symmetry breaking for all
operators, so that typically the corresponding primed and unprimed
coefficients are equal.  Values used for $Q_i$ are discussed in
Sec.~\ref{results} below.  Note that 3-body operators with 3 Pauli
matrices are absent~\cite{BHL} due to time-reversal symmetry.  The use
of primed and unprimed coefficients with the same labels echoes the
similar form of the operators in the two cases; indeed, in a number of
models the primed and unprimed coefficients are related~\cite{BG}.

One can obtain the most general charge radius and quadrupole moment
expressions by computing the 12 fundamental matrix elements
\begin{eqnarray}
\langle \alpha \beta \rangle^{(0)} & \equiv & \left< \xbf{S}_\alpha \!
\cdot \! \xbf{S}_\beta \right> , \nonumber \\
\langle \alpha \beta \rangle^{(2)} & \equiv & \left< 3 S_\alpha^z
S_\beta^z - \xbf{S}_\alpha \! \cdot \! \xbf{S}_\beta \right> ,
\label{ab2def}
\end{eqnarray}
where $(\alpha,\beta)$ are the 6 distinct pairs $(u,u)$, $(u,d)$,
$(u,s)$, $(d,d)$, $(d,s)$, and $(s,s)$ (Note the symmetry under
$\alpha \leftrightarrow \beta$).  Then, charge radii or quadrupole
moments for baryons in the presence of arbitrary SU(3) or isospin
symmetry breaking can be computed simply by taking an appropriate
linear combination of these primitive matrix elements.

In particular, while the indices $i,j,k$ indicate each of the $N_c$
quarks within the baryon, let the indices $\alpha, \beta, \gamma$
indicate collectively the quarks of a particular flavor $u, d$, or
$s$.  The coefficients for the most general operators that appear for
the charge radii or quadrupole moments can be represented by 3-vectors
such as $\xbf{a} = ( a_u, a_d, a_s)$.  Thus, matrix elements of the
1-body operator for charge radii generalize to:
\begin{equation}
\label{lin1}
\sum_i^{N_c} \, a_i = \sum_\alpha N_\alpha \, a_\alpha ,
\end{equation}
while those of the 2-body operator may be expanded as
\begin{equation} \label{cr1}
\left< \sum_{i \neq j} \, b_i \, c_j \; \xbf{\sigma}_i \! \cdot \!
\xbf{\sigma_j} \right> = 4 \sum_{\alpha , \beta} \, b_\alpha \,
c_\beta \; \langle \alpha \beta \rangle^{(0)} - 3 \sum_\alpha
N_\alpha \, b_\alpha \, c_\alpha ,
\end{equation}
and those of the 3-body operator become:
\begin{eqnarray} \label{cr2}
\left< \sum_{i \neq j \neq k} d_i \, e_j \, f_k \; \xbf{\sigma}_i \!
\cdot \! \xbf{\sigma}_j \right> & = & + 4 \sum_{\alpha , \beta}
d_\alpha \, e_\beta \; \langle \alpha \beta \rangle^{(0)} \;
\sum_\gamma N_\gamma f_\gamma \; \nonumber \\
& & - 4 \sum_{\alpha, \beta} d_\alpha \, e_\beta \, (f_\alpha +
f_\beta) \; \langle \alpha \beta \rangle^{(0)} \nonumber \\ & & - 3
\sum_\alpha N_\alpha \, d_\alpha \, e_\alpha \sum_\beta N_\beta \,
f_\beta \; + 6 \sum_\alpha N_\alpha \, d_\alpha \, e_\alpha
\, f_\alpha ,
\end{eqnarray}
where $N_\alpha$ is the number of quarks of type $\alpha$ (values
given in Table~\ref{ab0}), which satisfies the constraints $N_u + N_d
+ N_s = N_c$ and $N_u - N_d = 2I_3$.

Matrix elements of the 2- and 3-body operators for the quadrupole
moments expand as:
\begin{equation} \label{qm1}
\left< \sum_{i \neq j} \, b^\prime_i \, c^\prime_j \; ( 3 \sigma_{iz}
\sigma_{jz} - \xbf{\sigma}_i \! \cdot \! \xbf{\sigma}_j ) \right> = 4
\sum_{\alpha, \beta} \, b^\prime_\alpha \, c^\prime_\beta \; \langle
\alpha \beta \rangle^{(2)} ,
\end{equation}
and
\begin{eqnarray}
\left< \sum_{i \neq j \neq k} \, d^\prime_i \, e^\prime_j \,
f^\prime_k \; ( 3 \sigma_{iz} \sigma_{jz} - \xbf{\sigma}_i \! \cdot \!
\xbf{\sigma}_j ) \right> & = & + 4 \sum_{\beta, \gamma} \,
d^\prime_\beta \, e^\prime_\gamma
\; \langle \beta \gamma \rangle^{(2)} \; \sum_{\alpha} N_\alpha
f^\prime_\alpha \nonumber \\ & & - 4 \sum_{\alpha , \beta}
d^\prime_\alpha \, e^\prime_\beta \, (f^\prime_\alpha +
f^\prime_\beta)
\; \langle \alpha \beta \rangle^{(2)} .
\label{qm2}
\end{eqnarray}
The two examples of SU(3) symmetry breaking in the quadrupole
operators considered in the $N_c=3$ work of Ref.~\cite{BH2} correspond
to setting $b^\prime_\alpha = Q_\alpha (m/m_\alpha)^n$,
$c^\prime_\alpha = e^\prime_\alpha = m/m_\alpha$, $d^\prime_\alpha =
(m/m_\alpha)^n$, and $f^\prime_\alpha = Q_\alpha$, where $m_u = m_d
\equiv m$ and $m/m_s \equiv r$ represent SU(3) symmetry breaking
arising from different quark masses.  The quadratic (cubic) case in
\cite{BH2} is $n=1(2)$.

\subsection{Reduction of the Basis}

Even with this decomposition, it is not necessary to compute the
entire set of 12 matrix elements separately: They are related by a
number of simple constraints.  First, note that Eq.~(\ref{ket}) is
sensitive to the exchange of $u$ and $d$ quarks only through the
second CG coefficient, and the factor obtained through this exchange
is just $(-1)^{S_u + S_d - I}$.  Of course, the values of $S_\alpha$,
which simply count one-half the number of quarks of flavor $\alpha$ in
these baryons, remain unchanged from initial to final state.  The same
is true for $I_3 = S_u - S_d$, but the total isospin may change to a
value $I^\prime$.  One thus finds for an operator ${\cal O}$ that
\begin{equation} \label{inversion}
\langle I^\prime I_3 \left| {\cal O} (u \leftrightarrow d) \right| I
I_3 \rangle = (-1)^{I^\prime - I} \langle I^\prime - \! I_3 \left|
{\cal O} \right| I - \! I_3 \rangle .
\end{equation}

There are almost enough constraints to allow extraction of all the
matrix elements $\langle \alpha \beta \rangle^{(0)}$ in terms of
simple combinations of eigenvalues of the compatible operators of the
system.  To be precise, these are $\xbf{J}^2 \rightarrow J(J+1)$,
$J_z$, $\xbf{S}_\alpha^2 \rightarrow (N_\alpha/2) (N_\alpha /2 + 1)$
for $\alpha = u,d,s$, and $\xbf{I}^2 =
\xbf{S}_{ud}^2 = I(I+1)$.  Note that the eigenvalue $I_3 = S_u - S_d$
is not independent.  Clearly, $\langle \alpha \alpha \rangle^{(0)} =
(N_\alpha/2) (N_\alpha /2 + 1)$.  The other two constraints are:
\begin{eqnarray}
J(J+1) = \langle \xbf{J}^2 \rangle = \langle ( \xbf{S}_u + \xbf{S}_d +
\xbf{S}_s )^2 \rangle & =  & \sum_{\alpha, \beta} \langle \alpha \beta
\rangle^{(0)} , \nonumber \\
I(I+1) = \langle \xbf{S}_{ud}^2 \rangle = \langle ( \xbf{S}_u +
\xbf{S}_d )^2 \rangle & = & \langle u u \rangle^{(0)} + 2 \langle u d
\rangle^{(0)} + \langle d d \rangle^{(0)} . \label{constraint0}
\end{eqnarray}
From these constraints it is clear that only one of the two matrix
elements $\langle u s \rangle^{(0)}$ or $\langle d s \rangle^{(0)}$
need be computed directly from Eq.~(\ref{ket}), while the other can be
obtained either from the inversion rule Eq.~(\ref{inversion}) or from
the constraints Eqs.~(\ref{constraint0}).

As for the matrix elements $\langle \alpha \beta \rangle^{(2)}$, the
most obvious constraint reads:
\begin{equation} \label{cons2a}
\left< 3 J_z^2 - \xbf{J}^2 \right> = \left< 3( S_u^z + S_d^z +
S_s^z )^2 - ( \xbf{S}_u + \xbf{S}_d + \xbf{S}_s )^2 \right> =
\sum_{\alpha, \beta} \langle \alpha \beta \rangle^{(2)}.
\end{equation}
The l.h.s.\ of this expression is of course simple to compute; in the
stretched state ($J_z = J$) in which quadrupole moments are computed,
it equals $J(2J-1)$ for diagonal matrix elements and vanishes for
transitions.

Another constraint may be obtained by considering the combination:
\begin{equation} \label{cons2b}
\langle 3 (S_{ud}^z)^2 - \xbf{S}_{ud}^2 \rangle = \langle 3 (J_z -
S_s^z)^2 - \xbf{I}^2 \rangle = 3J_z^2 - 6J_z \langle S_s^z \rangle +
\langle s s \rangle^{(2)} + \langle \xbf{S}_s^2 \rangle - \langle
\xbf{I}^2 \rangle ,
\end{equation}
whose l.h.s.\ is just
\begin{equation} \label{cons2c}
\langle u u \rangle^{(2)} + 2 \langle u d \rangle^{(2)} + \langle d d
\rangle^{(2)} .
\end{equation}
Note that this constraint requires one to compute also the matrix
element $\langle S_s^z \rangle$, but this calculation turns out in
purely algebraic terms to be simpler than that of the rank-2 tensors.
One still requires one more constraint to compute separate values for
$\langle u s \rangle^{(2)}$ and $\langle d s \rangle^{(2)}$; once one
of these matrix elements is in hand, the other is obtained by using
Eq.~(\ref{inversion}).  The required constraint, again using $\langle
\xbf{S}_{ud}^2 \rangle = \langle \xbf{I}^2 \rangle$, may be obtained from:
\begin{equation} \label{cons2d}
\langle 3 S_u^z J_z - \xbf{S}_u \! \cdot \! \xbf{J} \rangle = 3 \langle
S_u^z \rangle J_z - \langle u s \rangle^{(0)} - \frac 1 2 \langle
\xbf{I}^2 + \xbf{S}_u^2 - \xbf{S}_d^2 \rangle ,
\end{equation}
where the l.h.s.\ clearly equals:
\begin{equation} \label{cons2e}
\langle u u \rangle^{(2)} + \langle u d \rangle^{(2)} + \langle u s
\rangle^{(2)} .
\end{equation}
In summary, then, the charge radii matrix elements can be computed in
complete generality using only eigenvalues and the calculation of
$\langle u s \rangle^{(0)}$, while the most general quadrupole moment
matrix elements can be obtained from further explicitly calculating
$\langle S_u^z \rangle$, $\langle S_s^z \rangle$, $\langle u u
\rangle^{(2)}$, and $\langle s s \rangle^{(2)}$.

Another advantage of using the
constraints~(\ref{constraint0})--(\ref{cons2d}) is that, by judicious
choice of which matrix elements to compute explicitly, one may obtain
results containing nothing more complicated than a $6j$ symbol.  These
appear due to the coupling of the quantum numbers of expansions for
both the bra and ket [Eq.~(\ref{ket})] through the operator.  For
example, in the matrix elements of $\langle u u \rangle^{(2)}$ defined
by Eq.~(\ref{ab2def}), one uses the Wigner-Eckart theorem to obtain:
\begin{equation}
\left< S_u S_u^z \left| \, 3 (S_u^z)^2 - \xbf{S}_u^2 \, \right|
S_u^\prime S_u^{\prime \, z} \right> = \delta_{S^\prime_u S_u}
\delta_{S^{\prime \, z}_u S^z_u} \sqrt{S_u (S_u+1) (2S_u-1)
(2S_u+3)} \left( \begin{array}{cc} 2 & S_u \\ 0 & S_u^z \end{array}
\right| \left. \begin{array}{c} S_u \\ S_u^z \end{array} \right) ,
\end{equation}
and the CG coefficient in this expression is linked to the those in
the bra and ket of Eq.~(\ref{ket}) by sums over $S_u^z$ and
$S_u^{\prime \, z}$.  Four appropriately linked CG coefficients
produce a $6j$ symbol, and six produce a $9j$ symbol.  In our case,
the matrix elements $\langle u s
\rangle^{(2)}$ and $\langle d s \rangle^{(2)}$ directly computed would
produce $9j$ symbols. In fact, so do $\langle u s \rangle^{(0)}$ and
$\langle d s \rangle^{(0)}$, but they produce a $9j$ symbol with one
zero argument, which can be written as the product of two $6j$
symbols, as seen below.  Analytic forms for $6j$ symbols with one
argument $\leq 2$ appear in Edmonds~\cite{edmonds}, which is precisely
what is needed to compute the matrix elements of tensors up to rank 2.

The analytic forms of the matrix elements of interest are:
\begin{eqnarray}
\langle u s \rangle^{(0)} & = &
\delta_{S^\prime S} \delta_{S_z^\prime S_z}
\delta_{S_u^\prime S_u} \delta_{S_d^\prime S_d} \delta_{S_s^\prime S_s}
(-1)^{1+S+S_s-S_u-S_d} \nonumber \\ & & \times
\sqrt{S_u (S_u+1) (2S_u+1) S_s (S_s + 1) (2S_s +1) (2I^\prime+1) (2I+1)}
\nonumber \\ & & \times
\sixj{S_d}{S_u}{I}{1}{I^\prime}{S_u} \sixj{S}{S_s}{I}{1}{I^\prime}{S_s} ,
\end{eqnarray}
\begin{eqnarray}
\langle S_u^z \rangle & = & \delta_{S_z^\prime S_z}
\delta_{S_u^\prime S_u} \delta_{S_d^\prime S_d} \delta_{S_s^\prime S_s}
(-1)^{S-S^\prime+S_z+S_s+I^\prime-I-S_u-S_d} \nonumber \\ & & \times
\sqrt{S_u (S_u+1) (2S_u+1) (2I^\prime+1) (2I+1) (2S^\prime + 1) (2S+1)}
\nonumber \\ & & \times \sixj{S_d}{S_u}{I}{1}{I^\prime}{S_u}
\sixj{S_s}{I}{S}{1}{S^\prime}{I^\prime}
\threej{1}{S^\prime}{S}{0}{S_z}{-S_z},
\end{eqnarray} 
\begin{eqnarray}
\langle S_s^z \rangle & = & \delta_{S_z^\prime S_z} \delta_{I^\prime I}
\delta_{S_u^\prime S_u} \delta_{S_d^\prime S_d} \delta_{S_s^\prime S_s}
(-1)^{1+S_z+S_s+I} \sqrt{S_s (S_s+1) (2S_s+1) (2S^\prime + 1) (2S+1)}
\nonumber \\ & & \times \sixj{I}{S_s}{S}{1}{S^\prime}{S_s}
\threej{1}{S^\prime}{S}{0}{S_z}{-S_z} ,
\end{eqnarray}
\begin{eqnarray}
\langle u u \rangle^{(2)} & = & \delta_{S_z^\prime S_z} \delta_{I^\prime I}
\delta_{S_u^\prime S_u} \delta_{S_d^\prime S_d} \delta_{S_s^\prime S_s}
(-1)^{S-S^\prime+S_z+S_s+I^\prime-I-S_u-S_d} \nonumber \\ & & \times
\sqrt{(2S_u-1) S_u(S_u+1)(2S_u+3)(2I^\prime+1)(2I+1)(2S^\prime+1)(2S+1)}
\nonumber \\ & & \times \sixj{S_d}{S_u}{I}{2}{I^\prime}{S_u}
\sixj{S_s}{I}{S}{2}{S^\prime}{I^\prime}
\threej{2}{S^\prime}{S}{0}{S_z}{-S_z} ,
\end{eqnarray}
\begin{eqnarray}
\langle s s \rangle^{(2)} & = & \delta_{S_z^\prime S_z} \delta_{I^\prime I}
\delta_{S_u^\prime S_u} \delta_{S_d^\prime S_d} \delta_{S_s^\prime S_s}
(-1)^{S_z+S_s+I} \nonumber \\ & & \times
\sqrt{(2S_s-1) S_s (S_s+1) (2S_s+1) (2S_s+3) (2S^\prime + 1) (2S+1)}
\nonumber \\ & & \times \sixj{I}{S_s}{S}{2}{S^\prime}{S_s}
\threej{2}{S^\prime}{S}{0}{S_z}{-S_z} .
\end{eqnarray}
Note that, in the interest of exhibiting maximal symmetry, the
remaining CG coefficients have been written as $3j$
symbols~\cite{edmonds}.

Charge radius and quadrupole transitions are diagonal in both charge
and strangeness, and thus connect only states of the same values of
$I_3$ and $N_s$, but do not necessarily conserve isospin.
Furthermore, charge radius operators are scalars and thus connect only
states of the same total spin $J$, but quadrupole operators are rank 2
and therefore can connect spin 3/2 to 3/2 or 1/2, but not 1/2 to 1/2.
These selection rules are reflected by the transition matrix elements
represented in the tables.  Values at arbitrary $N_c$ for the matrix
elements $\langle \alpha \beta \rangle^{(0)}$ ($\alpha \neq \beta$)
for all relevant states are collected in Table~\ref{ab0}, for
$S_\alpha^z$ in Table~\ref{S1}, for $\langle \alpha \alpha
\rangle^{(2)}$ in Table~\ref{aa2}, and $\langle \alpha \beta
\rangle^{(2)}$ ($\alpha
\neq \beta$) in Table~\ref{ab2}.  One may obtain results for charge
radii and quadrupole moments, including arbitrary SU(3) symmetry
breaking, by combining the results of these tables using the
expressions (\ref{lin1})--(\ref{qm2}) derived in the previous
Subsection.

\section{Results} \label{results}

We exhibit in the Appendices expressions for the baryon charge radii
(A) and quadrupole moments (B) under three sets of assumptions
familiar to researchers in the quark model.  The first is simply to
assume no SU(3) flavor symmetry breaking except for that inherent from
the quark charges.  The second and third, which we call ``quadratic''
($n=2$) and ``cubic'' ($n=3$) SU(3) symmetry breaking, respectively,
correspond to modifying the spin-spin terms in the following way:
\begin{equation} \label{quark}
\xbf{\sigma}_i \xbf{\sigma}_j \to \xbf{\sigma}_i \xbf{\sigma}_j \,
m^n /m_i^{n-1} m_j ,
\end{equation}
where $m_i$ denotes the constituent mass of quark $i$, and $m = m_u =
m_d$ is the light quark constituent mass, thus fixing all parameters
$a$-$f$ and $b^\prime$-$f^\prime$ defined in (\ref{chradop}) and
(\ref{quadop}).  Let us henceforth abbreviate with $m/m_s \equiv r$
the degree of SU(3) flavor symmetry breaking.  The quadratic mass
dependence arises in a constituent quark model with the dominant
interaction mechanism being one-gluon exchange, while the extra mass
factor for cubic mass dependence arises from a quark propagator
between photon absorption and gluon emission.

One can also mix these pictures so that, for example, the operator
labeled by $B(B^\prime)$ uses cubic SU(3) symmetry breaking, while
that labeled by $C(C^\prime)$ uses quadratic breaking.  Note that no
SU(3) symmetry breaking has been introduced in the one-body ($A$)
operator.

We hasten to add that it is not necessary to adopt {\em any\/} model
dependence beyond the single-photon exchange ansatz, since we have
computed and tabulated all the relevant primitive matrix elements.
However, the SU(3)-symmetric, quadratic, and cubic models provide a
useful picture in which to investigate the consequences of the
arbitrary $N_c$ calculation, without adding any new parameters.
Another interesting scheme for SU(3) symmetry breaking, but not
investigated here, is that provided by chiral perturbation theory.  In
this case, loop graphs that include the octet of light mesons as
Goldstone bosons produce terms with different $r$ dependences, {\it
e.g.}, logarithmic.

We have also used expressions~\cite{Shrock} for the quark charges that
simultaneously guarantee chiral anomaly cancellation of the
$N_c$-extended standard model and fix the total charges of the
$N_c$-quark baryons to equal their $N_c=3$ values.  These are:
\begin{equation} \label{qchg}
Q_{u,c,t} = (N_c+1)/2N_c , \ \ Q_{d,s,b} = (-N_c+1)/2N_c .
\end{equation}
Given the primitive matrix elements of Tables~\ref{ab0}--\ref{ab2},
one may alternately compute expressions using the strict $N_c=3$
values for the quark charges, with the caveats that the anomaly
cancellation conditions are no longer satisfied and the baryon charges
become $N_c$ dependent.

One may choose any of a number of schemes for obtaining interesting
predictions from our results.  Rather than selecting just one and
producing exhaustive results, we discuss several possibilities and
exhibit illustrative examples.

First, one must differentiate predictions that hold in the physical
case $N_c=3$ from those that hold in the $1/N_c \to 0$ limit.  The
former have the advantage of not depending on the quality of the
$1/N_c$ expansion at $N_c=3$, but may depend upon delicate
cancellations between terms at different powers in $1/N_c$.  Thus,
such predictions may hold well only for $N_c=3$ but not for
5,7,\ldots.  Conversely, the latter have the advantage of holding to a
desired level of accuracy for small values of $1/N_c$, but these
corrections may turn out to be numerically significant for $N_c=3$.
As we exhibit below, a number of relations are found to hold
independent of $N_c$ and thus satisfy both criteria.  Moreover, there
are even relations that hold both for $N_c=3$ and $1/N_c \to 0$, but
differ in between; we discuss one such example with the quadrupole
moments below.

\subsection{Charge Radii}

Baryon mean-square charge radii, denoted here by $r_B^2$, are defined
by the l.h.s.\ of Eq.~(\ref{chradop}), divided by the total charge if
it is nonzero.  In all suggested substitutions for the coefficients
$a_\alpha, \ldots , f_\alpha$, isospin violation is introduced via the
single-photon exchange ansatz through a single power of the quark
charge operator $Q_\alpha$, which transforms as a combination of $I=0$
and $I=1$.  Consequently, any combination of charge radii only
sensitive to $I=2$ or $I=3$ operators must vanish.  These are
\begin{eqnarray}
\lefteqn{\underline{I=2}:} & & \nonumber \\
0 & = & 2 r^2_{\Delta^{++}} - 3 r^2_{\Delta^+} + 3 r^2_{\Delta^0} +
r^2_{\Delta^-} \label{D1} , \\
0 & = & r^2_{\Sigma^+} - 2 r^2_{\Sigma^0} - r^2_{\Sigma^-} ,
\label{Sig1} \\
0 & = & r^2_{\Sigma^{*+}} - 2 r^2_{\Sigma^{*0}} - r^2_{\Sigma^{*-}} ,
\label{Sig2} \\
\lefteqn{\underline{I=3}:} & & \nonumber \\
0 & = & 2 r^2_{\Delta^{++}} - r^2_{\Delta^+} - r^2_{\Delta^0} -
r^2_{\Delta^-} . \label{D2}
\end{eqnarray}
Equations~(\ref{D1}) and (\ref{D2}) were first derived in
Ref.~\cite{BL1}, while Eqs.~(\ref{Sig1}) and (\ref{Sig2}) are $\Sigma$
equal-spacing rules, adjusted for the negative charge of
$\Sigma^{(*)-}$.  In addition, using Appendix~A one finds 3 linear
combinations of charge radii with $N_c$-independent coefficients that
vanish for arbitrary values of $N_c$ and $r$, either in the quadratic
or cubic case of SU(3) symmetry breaking:
\begin{eqnarray}
0 & = & -20(r^2_p + r^2_n) + 5(r^2_{\Sigma^+} - r^2_{\Sigma^0} +
3r^2_{\Sigma^-}) + 5r^2_\Lambda - 4 ( 4r^2_{\Delta^{++}} -
r^2_{\Delta^+} - 4r^2_{\Delta^0} + 7r^2_{\Delta^-} ) \nonumber \\ & &
+ 10 (r^2_{\Sigma^{*+}} - r^2_{\Sigma^{*0}} + 3r^2_{\Sigma^{*-}}) , \\
0 & = & 4(r^2_p - 5 r^2_n) - (5r^2_{\Sigma^+} +3 r^2_{\Sigma^0} -
r^2_{\Sigma^-}) + 35r^2_\Lambda +4 ( 2r^2_{\Delta^{++}} +
r^2_{\Delta^+} +r^2_{\Delta^0} -r^2_{\Delta^-} ) \nonumber \\ & &
-2 (5r^2_{\Sigma^{*+}} +3r^2_{\Sigma^{*0}} - r^2_{\Sigma^{*-}}) , \\
0 & = & -8(r^2_p + r^2_n) + (r^2_{\Sigma^+} \! - 9r^2_{\Sigma^-} \! )
+ 2r^2_\Lambda \! + 16 r^2_{\Xi^-} \! + 2(r^2_{\Sigma^{*+}} \! -
9r^2_{\Sigma^{*-}} \! ) + 32 r^2_{\Xi^{*-}} \! - 16 r^2_{\Omega^-} .
\end{eqnarray}
Since so many of these charge radii are currently unmeasured, these
relations are presented merely for completeness.

A number of charge radii can be related if one permits $N_c$-dependent
coefficients.  A particularly pretty example is
\begin{equation} \label{rnlam}
r^2_\Lambda = \frac{3}{N_c+3} r_n^2 .
\end{equation}
The $N_c=3$ version of this relation, but applied to magnetic moments
rather than charge radii, is known from the early days of SU(3) flavor
(see, {\it e.g.}, \cite{Cole}).  Using the measured value for the
neutron charge radius~(\ref{rnmeas}) and $N_c=3$, one predicts
$r^2_\Lambda = -0.057(3)$ fm$^2$.  The expression (\ref{rnlam}) is
especially interesting because it definitively predicts---independent
of SU(3) symmetry breaking---that the sign of the charge radius of
$\Lambda$ is the same as that of the neutron, {\it i.e.}, negative.
This is not the case, for example, in the calculations of
Ref.~\cite{Thomas} (extrapolation from lattice results) or
Ref.~\cite{WBF} (constituent quark model).  For other comparisons,
Refs.~\cite{WBF,Povh,Sahoo} also contain tabulated results of various
authors' baryon charge radii calculations.

Another interesting result is that
\begin{equation}
r^2_{\Sigma^+} = r^2_p + O(1-r),
\end{equation}
so that the two charge radii are equal in the SU(3) limit {\em for all
values of\/} $N_c$, as is evident from Eqs.~(\ref{crp}) and
(\ref{crsp}).  One expects that this relationship holds to better
accuracy than $r^2_{\Xi^0} = r_n^2$, which requires {\em both\/} the
SU(3) limit and $N_c=3$ to hold, so that both corrections of $O(1-r)$
and $O(1-3/N_c)$ occur.

There is a relation of the latter type between the 3 measured charge
radii:
\begin{equation} \label{rpns}
r^2_p - r^2_{\Sigma^-} + r_n^2 = O(1-r) + O(1-3/N_c) .
\end{equation}
Or, if one allows $N_c$-dependent coefficients,
\begin{equation} \label{ncdep}
\frac 1 2 (N_c-1) (r^2_p - r^2_{\Sigma^-}) + r^2_n = O[(1-r)(1-3/N_c)]
.
\end{equation}
Numerically, the l.h.s.\ nearly vanishes: One finds $-0.06(15)$
fm$^2$.  However, the uncertainty is the figure of merit here, since
it may be used to gauge the typical size of a first-order
SU(3)-breaking effect.

Yet one more interesting result is:
\begin{equation}
r^2_{\Xi^-} = \frac 1 2 \left( r^2_p + r^2_{\Sigma^-} + r_n^2 \right)
+ O[(1-r)^2] + O (1-3/N_c) ,
\end{equation}
in both the quadratic and cubic forms of SU(3) symmetry breaking.  The
second of the order terms implies that the result holds exactly only
for $N_c=3$, but this can be pushed to higher order if one allows
$N_c$-dependent coefficients on the l.h.s., just as in
Eq.~(\ref{ncdep}).  The virtue of this expression as it stands is that
it predicts $r^2_{\Xi^-}$ up to second order in SU(3) symmetry
breaking entirely in terms of the known charge radii, with simple
coefficients.  One finds
\begin{equation}
r^2_{\Xi^-} = 0.64(8) \ {\rm fm}^2 ,
\end{equation}
where the uncertainty is dominated by that of the $r^2_{\Sigma^-}$
measurement, which is much larger than the piece of the uncertainty
one would estimate from the second-order SU(3) symmetry breaking.

Lastly, our expressions for charge radii depend upon only four
parameters, $A,B,C,$ and $r$ [and a choice of scheme for SU(3)
breaking].  Once one additional charge radius is measured, it will be
posssible to solve for all the parameters and predict all of the other
charge radii.  In this sense, $r^2_{\Xi^-}$ was chosen in the previous
paragraph because it could be predicted in terms of just the measured
charge radii, with little sensitivity to the SU(3)-breaking scheme or
the parameter $r$.  Conversely, if it becomes the fourth measured baryon
charge radius, it will not effectively constrain the parameter $r$.

On the other hand, it turns out that the value of $r^2_{\Omega^-}$ (to
give one example) cannot be predicted in terms of $r^2_{p, \, n,
\, \Sigma^-}$, even in the SU(3) limit.  Turning this unfortunate
observation around means that a measurement of $r^2_{\Omega^-}$ would
provide an extremely sensitive probe of SU(3) symmetry breaking, and
permit high-quality predictions of all other charge radii.

Using the three measured charge radii and the assumption of either
quadratic ($n=1$) or cubic ($n=2$) SU(3) symmetry breaking and an
assumed value of $r$, one can obtain at least ranges for the values of
the parameters $A$, $B$, and $C$.  Since these parameters have an
expansion in $1/N_c^m$ starting with $m=0$, the values obtained for
$N_c=3$ hold for arbitrary $N_c$.  One finds:
\begin{eqnarray}
(1-2r-2r^n)(r^2_p + r^2_n) +3 r^2_{\Sigma^-} & = & +2(2-r-r^n) A ,
\nonumber \\
r^2_p - r^2_{\Sigma^-} + \frac 1 3 (5-r-r^n) r^2_n & = & -\frac 2 3 (2
- r - r^n) B , \nonumber \\
r^2_p - r^2_{\Sigma^-} - \frac 1 3 (1-2r-2r^n) r^2_n & = & -\frac 4 9
(2 - r - r^n) C .
\end{eqnarray}
Note that these all become Eq.~(\ref{rpns}) in the SU(3) limit.  The
uncertainties in these expressions are dominated by that of
$r^2_{\Sigma^-}$, even more than by that of $n$.  Including only the
former and assuming the value $r=$ 330 MeV$\!$/$\,$540 MeV suggested
by the constituent quark model, one computes:
\begin{equation}
A = +0.56 \pm 0.29 \, {\rm fm}^2, \
B = -0.05 \pm 0.29 \, {\rm fm}^2, \
C = -0.33 \pm 0.44 \, {\rm fm}^2 .
\end{equation}
With these central values one predicts $r^2_{\Delta^+} < r^2_p$ for
$N_c=3$, contrary to the physical picture that the $\Delta$ is an
excited state of the nucleon and hence is more extended in space.
However, it is well within the error bars of $r^2_{\Sigma^-}$ for $B$
to be substantial and for $C$ to nearly vanish. For example, with
$r^2_{\Sigma^-}$=0.73 fm$^2$ as suggested by the upper value of the
range of statistical uncertainty in Eq.~(\ref{rsigmes}) and $r$ as
above one obtains $A=0.79$, $B=0.18$, and $C=0.02$, a hierachy of
parameters leading to the reasonable conclusion $r^2_{\Delta^+} >
r^2_p$.

\subsection{Quadrupole Moments}

The operator (\ref{quadop}) defining quadrupole moments shares with
the charge radius operator (\ref{chradop}) the feature in the
single-photon exchange ansatz that isospin violation enters only
through a single power of the quark charge operator $Q_\alpha$.  Thus,
as before, combinations sensitive only to $I=2$ or $I=3$ operators
must vanish.  These are:
\begin{eqnarray}
\lefteqn{\underline{I=3}:} & & \nonumber \\
0 & = & Q_{\Delta^{++}} - 3 Q_{\Delta^+} + 3 Q_{\Delta^0} -
Q_{\Delta^-} , \\
\lefteqn{\underline{I=2}:} & & \nonumber \\
0 & = & Q_{\Delta^{++}} - Q_{\Delta^+} - Q_{\Delta^0} + Q_{\Delta^-} ,
\\
0 & = & Q_{\Delta^+ p} - Q_{\Delta^0 n} , \\
0 & = & Q_{\Sigma^{*+}} - 2 Q_{\Sigma^{*0}} + Q_{\Sigma^{*-}} , \\
0 & = & Q_{\Sigma^{*+} \Sigma^+} - 2 Q_{\Sigma^{*0} \Sigma^0} +
Q_{\Sigma^{*-} \Sigma^-} .
\end{eqnarray}
The first three of these expressions were obtained in Ref.~\cite{BHL},
while the last two are $\Sigma^{(*)}$ equal-spacing rules, obtained
for $N_c=3$ in \cite{RLSU6} and \cite{BH2}.  In addition, there is
precisely one linear relation with $N_c$-independent coefficients that
holds for all values of $N_c$ in all cases of SU(3) symmetry breaking
studied here:
\begin{equation}
0 = Q_{\Xi^{*-}} - Q_{\Omega^-} - \sqrt{2} Q_{\Xi^{*-} \Xi^-} .
\end{equation}

Unlike the charge radius case, only the $N \to \Delta$ quadrupole
transition matrix element has been measured (via photoproduction
experiments~\cite{LEGS}), and even here the extraction of $Q_{N \to
\Delta}$ is plagued by a large model dependence:
\begin{equation}
Q_{N \to \Delta} = -0.108 \pm 0.009 \, ({\rm stat \, + \, syst})
\pm 0.034 \, ({\rm model}) \ {\rm fm}^2.
\end{equation}
Since there are 3 undetermined parameters ($B^\prime, C^\prime,$ and
$r$, as well as a choice of SU(3)-breaking scheme), we do not attempt
to predict any of the other quadrupole moments numerically.  However,
we can still make a number of interesting observations based on the
structure of the expressions in Appendix~B.

First note that each quadrupole moment expression is $O(N_c^0)$ or
smaller in the $1/N_c$ expansion.  Indeed, only one coefficient,
$B^\prime$, contributes to this leading order (in the case of charge
radii, both $A$ [for charged baryons only] and $B$ contribute at
leading order).  A few moments' study will confirm that the diagonal
quadrupole moments are given by the expression
\begin{equation}
Q (I_3, Y) = I_3 [1 - Y(4-Y)/15] B^\prime + O(1/N_c) .
\end{equation}
Proportionality to $I_3$ also holds for the leading terms of the
transition quadrupole moments when the initial and final baryon states
have the same value of $I$ ($\Sigma^* \Sigma$ and $\Xi^* \Xi$).  This
behavior, due to the dominance of the isovector portion of the
quadrupole operator, is familiar from Skyrme model and a variety of
other model calculations~\cite{Oh,CB}.  Taken at face value, it
predicts an appreciable quadrupole moment for $\Delta^0$, as well as
$Q_{\Omega^-} = 0$.  However, the subleading corrections in $1/N_c$,
which interpolate between the extreme $N_c \to \infty$ and $N_c = 3$
cases, soften this behavior.  In particular, the diagonal quadrupole
moments in the strict $N_c=3$ case with no SU(3) flavor symmetry
breaking obey
\begin{equation}
Q = 4q/3 \left( B^\prime + C^\prime/3 \right) ,
\end{equation}
where $q$ is the baryon charge.

As discussed in Sec.~\ref{ops}, the charge radius and quadrupole
operators are very similar in that both represent spin-dependent
electromagnetic couplings to baryons.  It thus should not be
surprising that their coefficients are related in a typical model.  In
Ref.~\cite{BHL} we saw that the one-gluon exchange picture gives rise
to the relation
\begin{equation} \label{Qrn}
Q_{\Delta^+ p} = \frac{1}{\sqrt{2}} r^2_n \frac{N_c}{N_c+3}
\sqrt{\frac{N_c+5}{N_c-1}} ,
\end{equation}
for which the factor on the r.h.s.\ following $r_n^2$ equals unity
both for $N_c = 3$ and $N_c \to \infty$.  In fact, this result can be
obtained in a much more general setting.  The key constraints needed
to obtain Eq.~(\ref{Qrn}) are $B^\prime = -2B$, $C^\prime = -2C$, and
as argued in Ref.~\cite{BG} for $N_c=3$, the same relation may be
derived using not only one-gluon exchange, but one-pion exchange or
scalar exchanges, or a mixture of these.

A particularly useful measurement for determining the coefficients in
the quadrupole sector would be that of $Q_{\Omega^-}$.  As we see in
Appendix~B, the value of $Q_{\Omega^-}$ is very sensitive to the
precise nature of SU(3) symmetry breaking, even more so than
$r^2_{\Omega^-}$.  We discussed in Ref.~\cite{BHL} ideas in the
literature for the experimental determination of $Q_{\Omega^-}$; while
such experiments are challenging, they appear to be feasible.  The
value of $Q_{\Omega^-}$ would teach us much about the shape of baryon
wave functions and the nature of SU(3) flavor symmetry.

\section{Conclusions} \label{concl}

We have presented techniques that permit the calculation of matrix
elements of operators acting upon 3-flavor baryon states with
arbtitrary $N_c$.  The approach requires only SU(2) Clebsch-Gordan
coefficients and combinations of them in the form of $6j$ symbols.  We
tabulated the values of a set of primitive operators for all relevant
states and demonstrated how they can be combined to give results for
charge radii or quadrupole moments, using any chosen pattern of SU(3)
(or isospin) flavor symmetry breaking.  In particular, we presented in
the Appendices results using the single-photon exchange ansatz,
augmented by either no SU(3) symmetry breaking, or one of two popular
types of SU(3) symmetry breaking suggested by (but not limited to) the
quark model with gluon exchange.  We obtained a large number of
interesting predictions and demonstrated how many others can be made
either working in some particular model, or once a small number of
additional baryon charge radii or quadrupole moments are
experimentally measured.

\section*{Acknowledgments}

A.J.B.\ thanks the Deutsche Forschungsgemeinschaft for some support
under title BU813/3-1.  R.F.L.\ thanks the U.S.\ Department of Energy
for support under Grant No.\ DE-AC05-84ER40150, and the High Energy
Theory Group of the University of California, San Diego for their
hospitality.

\appendix
\section{Charge Radius Expressions}

We use the subscripts $0$, $Q$, and $C$ to denote expressions with
zero, quadratic, and cubic SU(3) flavor symmetry breaking via
constituent quark masses [see Eq.~(\ref{quark})] in addition to that
provided by the quark charge operator.  In the SU(3) symmetry limit
($r=1$) the expressions with subscripts $Q$ and $C$ reduce to those
with subscript $0$.

\begin{eqnarray}
r^2_{0, \, Q, \, C} \, (\Delta^{++}) & = & A + B \,
\frac{3(N_c^2-2N_c+5)}{4N_c^2} - C \, \frac{3(3N_c^2-12 N_c+5)}
{2N_c^3} , \\
r^2_{0, \, Q, \, C} \, (\Delta^+) & = & A + B \,
\frac{N_c^2-4N_c+15}{2N_c^2} - C \, \frac{(N_c-1)(4N_c-15)}{N_c^3} , \\
r^2_{0, \, Q, \, C} \, (\Delta^0) & = & - \left( B - \frac{2C}{N_c}
\right) \frac{(N_c-3)(N_c-5)}{2N_c} , \\
r^2_{0, \, Q, \, C} \, (\Delta^-) & = & A + B \,
\frac{3(N_c^2-5)}{2N_c^2} - C \, \frac{3(2N_c^2-5N_c-5)}{N_c^3} , \\
r^2_0 \, (\Sigma^{*+}) & = & A + B \, \frac{5N_c^2-17N_c+30}{4N_c^2} -
C \, \frac{11N_c^2-47N_c+30}{2N_c^3} , \nonumber \\
r^2_Q \, (\Sigma^{*+}) & = & A + \frac B 4 \left[ 4 + r -
\frac{14+3r}{N_c} + \frac{2(11+4r)}{N_c^2} \right] \nonumber \\ & &
- \frac{C}{2N_c} \left[ 10+r - \frac{36+11r}{N_c} +
\frac{2(11+4r)}{N_c^2} \right] ,
\nonumber \\
r^2_C \, (\Sigma^{*+}) & = & A + \frac B 4 \, \left[ 4+r -
\frac{14-r+4r^2}{N_c} + \frac{2(11+2r+2r^2)}{N_c^2} \right] \nonumber
\\ & & - \frac{C}{4N_c} \left[ 20+r+r^2 - \frac{72+11r+11r^2}{N_c} +
\frac{4(11+2r+2r^2)}{N_c^2} \right] , \\
r^2_0 \, (\Sigma^{*0}) & = & - \left( B - \frac{2C}{N_c}
\right) \frac{5(N_c-3)}{2N_c^2} , \nonumber \\
r^2_Q \, (\Sigma^{*0}) & = & - \left( B - \frac{2C}{N_c} \right)
\frac{1}{2N_c} \left( 3+2r - \frac{11+4r}{N_c} \right) ,
\nonumber \\
r^2_C \, (\Sigma^{*0}) & = & - \frac{B}{2N_c} \, \left( 3+2r^2 -
\frac{11+2r+2r^2}{N_c} \right) + \frac{C}{N_c^2}
\left( 3+r+r^2 - \frac{11+2r+2r^2}{N_c} \right) , \nonumber \\ & & \\
r^2_0 \, (\Sigma^{*-}) & = & A + B \, \frac{5N_c^2+3N_c-30}{4N_c^2} -
C \, \frac{11N_c^2-27N_c-30}{2N_c^3} , \nonumber \\
r^2_Q \, (\Sigma^{*-}) & = & A + \frac B 4 \left[ 4+r -
\frac{2-5r}{N_c} - \frac{2(11+4r)}{N_c^2} \right] \nonumber \\ & &
- \frac{C}{2N_c} \left[ 10+r - \frac{3(8+r)}{N_c} -
\frac{2(11+4r)}{N_c^2} \right] ,
\nonumber \\
r^2_C \, (\Sigma^{*-}) & = & A + \frac B 4 \left[ 4+r -
\frac{2-r-4r^2}{N_c} - \frac{2(11+2r+2r^2)}{N_c^2} \right] \nonumber
\\ & & - \frac{C}{4N_c} \left[ 20+r+r^2 - \frac{3(16+r+r^2)}{N_c} -
\frac{4(11+2r+2r^2)}{N_c^2} \right] , \\
r^2_0 \, (\Xi^{*0}) & = & + \left( B - \frac{2C}{N_c}
\right) \frac{5(N_c-3)^2}{6N_c^2} , \nonumber \\
r^2_Q \, (\Xi^{*0}) & = & +\left( B - \frac{2C}{N_c} \right) \frac 1 6 
\left[ 3+2r - \frac{6(3+r+r^2)}{N_c} + \frac{3(9+4r+2r^2)}{N_c^2}
\right] , \nonumber \\
r^2_C \, (\Xi^{*0}) & = & +\frac B 6 \left[ 3+2r -
\frac{6(3+r^2+r^3)}{N_c} + \frac{3(9+2r+2r^2+2r^3)}{N_c^2} \right]
\nonumber \\ & & + \frac{C}{3N_c} \left[ 3+r+r^2 -
\frac{3(6+r+r^2+2r^3)}{N_c} + \frac{3(9+2r+2r^2+2r^3)}{N_c^2} \right]
, \\
r^2_0 \, (\Xi^{*-}) & = & A + B \, \frac{5N_c^2+12N_c-45}{6N_c^2} - C
\frac{14N_c^2-33N_c-45}{3N_c^3} , \nonumber
\\
r^2_Q \, (\Xi^{*-}) & = & A + \frac{B}{6} \left[ 3 + 2r +
\frac{6r(1+r)}{N_c} - \frac{3(9+4r+2r^2)}{N_c^2} \right] \nonumber
\\ & & - \frac{C}{3N_c} \left[ 2(6+r) - \frac{3(9+2r)}{N_c} -
\frac{3(9+4r+2r^2)}{N_c^2} \right] , \nonumber \\
r^2_C \, (\Xi^{*-}) & = & A + \frac{B}{6} \left[ 3+2r +
\frac{6r^2(1+r)}{N_c} - \frac{3(9+2r+2r^2+2r^3)}{N_c^2} \right] 
\nonumber \\ & & - \frac{C}{3N_c} \left[ 12+r+r^2 -
\frac{3(9+r+r^2)}{N_c} - \frac{3(9+2r+2r^2+2r^3)}{N_c^2} \right] , \\
r^2_0 \, (\Omega^-) & = & A + B \, \frac{3(3N_c-5)}{2N_c^2} - C \,
\frac{3(N_c^2-2N_c-5)}{N_c^3} , \nonumber \\
r^2_Q \, (\Omega^-) & = & A + \frac{3B}{2N_c} \left( 1+2r^2 -
\frac{3+2r^2}{N_c} \right) - \frac{3C}{N_c} \left( 1 - \frac{2}{N_c} -
\frac{3+2r^2}{N_c^2} \right) , \nonumber \\
r^2_C \, (\Omega^-) & = & A + \frac{3B}{2N_c} \left( 1+2r^3 -
\frac{3+2r^3}{N_c} \right) - \frac{3C}{N_c} \left( 1 - \frac{2}{N_c} -
\frac{3+2r^3}{N_c^2} \right) , \\
r^2_{0, \, Q, \, C} \, (p) & = & A + B \,
\frac{(N_c-3)(N_c-1)}{2N_c^2} - C \, \frac{(N_c-1)(4N_c-3)}{N_c^3} ,
\label{crp} \\
r^2_{0, \, Q, \, C} \, (n) & = & - \left( B - \frac{2C}{N_c} \right)
\frac{(N_c-1)(N_c+3)}{2N_c^2} , \\
r^2_0 \, (\Sigma^+) & = & A + B \, \frac{(N_c-3)(N_c-1)}{2N_c^2} - C
\, \frac{(N_c-1)(4N_c-3)}{N_c^3} , \nonumber \\
r^2_Q \, (\Sigma^+) & = & A + \frac{B}{2} \left( 2-r -
\frac{7-3r}{N_c} + \frac{11-8r}{N_c^2} \right) - \frac{C}{N_c} \left(
5-r - \frac{18-11r}{N_c} + \frac{11-8r}{N_c^2} \right) , \nonumber \\
r^2_C \, (\Sigma^+) & = & A + \frac{B}{2} \left( 2-r -
\frac{7+r-4r^2}{N_c} + \frac{11-4r-4r^2}{N_c^2} \right) \nonumber \\ &
& - \frac{C}{2N_c} \left[ 10-r-r^2 - \frac{36-11r-11r^2}{N_c} +
\frac{2(11-4r-4r^2)}{N_c^2} \right] , \label{crsp} \\
r^2_0 \, (\Sigma^0) & = & + \left( B - \frac{2C}{N_c} \right)
\frac{N_c+3}{2N_c^2} , \nonumber \\
r^2_Q \, (\Sigma^0) & = & - \left(B - \frac{2C}{N_c} \right)
\frac{1}{2N_c} \left( 3-4r -\frac{11-8r}{N_c} \right), \nonumber \\
r^2_C \, (\Sigma^0) & = & -\frac{B}{2N_c} \left( 3-4r^2
-\frac{11-4r-4r^2}{N_c} \right) + \frac{C}{N_c^2} \left( 3-2r-2r^2
-\frac{11-4r-4r^2}{N_c} \right) , \nonumber \\ & & \\
r^2_{0, \, Q, \, C} \, (\Lambda) & = & - \left( B - \frac{2C}{N_c}
\right) \frac{3(N_c-1)}{2N_c^2} , \\
r^2_0 \, (\Sigma^0 \Lambda) & = & -\left( B - \frac{2C}{N_c} \right)
\frac{\sqrt{(N_c-1)(N_c+3)}}{2N_c} ,
\nonumber \\
r^2_Q \, (\Sigma^0 \Lambda) & = & -r \left( B - \frac{2C}{N_c} \right)
\frac{\sqrt{(N_c-1)(N_c+3)}}{2N_c}, \nonumber \\
r^2_C \, (\Sigma^0 \Lambda) & = & -r \left[ B - \frac{C(1+r)}{N_c}
\right] \frac{\sqrt{(N_c-1)(N_c+3)}}{2N_c}, \\
r^2_0 \, (\Sigma^-) & = & A + B \, \frac{N_c^2-6N_c-3}{2N_c^2} - C \,
\frac{4N_c^2-9N_c-3}{N_c^3}, \nonumber \\
r^2_Q \, (\Sigma^-) & = & A + \frac B 2 \left( 2-r -
\frac{1+5r}{N_c} - \frac{11-8r}{N_c^2} \right) - \frac{C}{N_c^2}
\left[ 5-r -\frac{3(4-r)}{N_c} -\frac{11-8r}{N_c^2} \right] ,
\nonumber \\
r^2_C \, (\Sigma^-) & = & A + \frac B 2 \left( 2-r
-\frac{1+r+4r^2}{N_c} -\frac{11-4r-4r^2}{N_c^2} \right) \nonumber \\
& &  -\frac{C}{2N_c^2} \left[ 10-r-r^2 -\frac{3(8-r-r^2)}{N_c}
-\frac{2(11-4r-4r^2)}{N_c^2} \right] , \\
r^2_0 \, (\Xi^0) & = & -\left( B - \frac{2C}{N_c} \right)
\frac{N_c^2+12N_c-9}{6N_c^2} , \nonumber \\
r^2_Q \, (\Xi^0) & = & +\left( B - \frac{2C}{N_c} \right) \frac 1 6
\left[ 3-4r -\frac{6(3-2r+r^2)}{N_c} +\frac{3(9-8r+2r^2)}{N_c^2}
\right] , \nonumber \\
r^2_C \, (\Xi^0) & = & +\frac B 6 \left[ 3-4r
-\frac{6(1+r)(3-3r+r^2)}{N_c} +\frac{3(9-4r-4r^2+2r^3)}{N_c^2} \right]
\nonumber \\ & & -\frac{C}{3N_c} \left[ 3-2r-2r^2
-\frac{6(3-r-r^2+r^3)}{N_c} +\frac{3(9-4r-4r^2+2r^3)}{N_c^2} \right] ,
\\
r^2_0 \, (\Xi^-) & = & A - B \, \frac{(N_c+3)^2}{6N_c^2} - C \,
\frac{8N_c^2-15N_c-9}{3N_c^3} , \nonumber \\
r^2_Q \, (\Xi^-) & = & A + \frac B 6 \left[ 3-4r -\frac{6r(2-r)}{N_c}
-\frac{3(9-8r+2r^2)}{N_c^2} \right] \nonumber \\ & & -\frac{C}{3N_c}
\left[ 4(3-r) -\frac{3(9-4r)}{N_c} -\frac{3(9-8r+2r^2)}{N_c^2} \right]
, \nonumber \\
r^2_C \, (\Xi^-) & = & A + \frac B 6 \left[ 3-4r
-\frac{6r^2(2-r)}{N_c} -\frac{3(9-4r-4r^2+2r^3)}{N_c^2} \right]
\nonumber \\ & &  -\frac{C}{3N_c} \left[ 2(6-r-r^2)
-\frac{3(9-2r-2r^2)}{N_c} -\frac{3(9-4r-4r^2+2r^3)}{N_c^2} \right] .
\end{eqnarray}

\section{Quadrupole Moment Expressions}

We use the subscripts $0$, $Q$, and $C$ to denote expressions with
zero, quadratic, and cubic SU(3) breaking via constituent quark masses
[see Eq.~(\ref{quark})] in addition to that provided by the quark
charge operator.  In the SU(3) symmetry limit ($r=1$) the expressions
with subscripts $Q$ and $C$ reduce to those with subscript $0$.

\begin{eqnarray}
Q_{0, \, Q, \, C} \, (\Delta^{++}) & = & +B^\prime \,
\frac{6(N_c^2+2N_c+5)}{5N_c^2} - C^\prime \,
\frac{12(N_c^2-8N_c+5)}{5N_c^3} , \\
Q_{0, \, Q, \, C} \, (\Delta^+) & = & +B^\prime \,
\frac{2(N_c^2+2N_c+15)}{5N_c^2} - C^\prime \,
\frac{4(N_c^2-13N_c+15)}{5N_c^3} , \\
Q_{0, \, Q, \, C} \, (\Delta^0) & = & - \left( B^\prime -
\frac{2C^\prime}{N_c} \right) \frac{2(N_c-3)(N_c+5)}{5N_c^2} , \\
Q_{0, \, Q, \, C} \, (\Delta^-) & = & -B^\prime \,
\frac{6(N_c^2+2N_c-5)}{5N_c^2} + C^\prime \,
\frac{12(N_c^2-3N_c-5)}{5N_c^3}, \\
Q_0 \, (\Sigma^{*+}) & = & +B^\prime \, \frac{N_c^2-N_c+6}{N_c^2} -
C^\prime \, \frac{2(N_c-6)(N_c-1)}{N_c^3} , \nonumber \\
Q_Q \, (\Sigma^{*+}) & = & +\frac{B^\prime}{2} \left[ 1+r +
\frac{1-3r}{N_c} + \frac{4(1+2r)}{N_c^2} \right] -
\frac{C^\prime}{N_c} \left[ 1+r - \frac{3+11r}{N_c} +
\frac{4(1+2r)}{N_c^2} \right] , \nonumber \\
Q_C \, (\Sigma^{*+}) & = & +\frac{B^\prime}{2} \left[ 1+r +
\frac{1+r-4r^2}{N_c} + \frac{4(1+r+r^2)}{N_c^2} \right]
\nonumber \\ & & -\frac{C^\prime}{2N_c} \left[ 2+r+r^2 -
\frac{6+11r+11r^2}{N_c} + \frac{8(1+r+r^2)}{N_c^2} \right] , \\
Q_0 \, (\Sigma^{*0}) & = & -\left( B^\prime - \frac{2C^\prime}{N_c}
\right) \frac{2(N_c-3)}{N_c^2} , \nonumber \\
Q_Q \, (\Sigma^{*0}) & = &  -\left( B^\prime - \frac{2C^\prime}{N_c}
\right) \frac{2}{N_c} \left( r - \frac{1+2r}{N_c} \right) ,
\nonumber \\
Q_C \, (\Sigma^{*0}) & = & -\frac{2B^\prime}{N_c} \left(r^2 -
\frac{1+r+r^2}{N_c} \right) + \frac{2C^\prime}{N_c^2}
\left[ r(1+r) - \frac{2(1+r+r^2)}{N_c} \right] , \\
Q_0 \, (\Sigma^{*-}) & = & -B^\prime \, \frac{N_c^2+3N_c-6}{N_c^2} +
C^\prime \, \frac{2(N_c^2-3N_c-6)}{N_c^3} , \nonumber \\
Q_Q \, (\Sigma^{*-}) & = & -\frac{B^\prime}{2} \left[ 1+r +
\frac{1+5r}{N_c} - \frac{4(1+2r)}{N_c^2} \right] +
\frac{C^\prime}{N_c} \left[ 1+r -\frac{3(1+r)}{N_c} -
\frac{4(1+2r)}{N_c^2} \right] , \nonumber \\
Q_C \, (\Sigma^{*-}) & = & -\frac{B^\prime}{2} \left[ 1+r
+\frac{1+r+4r^2}{N_c} - \frac{4(1+r+r^2)}{N_c^2} \right] \nonumber \\
& & + \frac{C^\prime}{2N_c} \left[ 2+r+r^2 -\frac{3(2+r+r^2)}{N_c} -
\frac{8(1+r+r^2)}{N_c^2} \right] , \\
Q_0 \, (\Xi^{*0}) & = & + \left( B^\prime - \frac{2C^\prime}{N_c}
\right) \frac{2(N_c-3)^2}{3N_c^2} , \nonumber \\
Q_Q \, (\Xi^{*0}) & = & + \left( B^\prime - \frac{2C^\prime}{N_c}
\right) \frac{2r}{3} \left[ 1 -\frac{3(1+r)}{N_c} +
\frac{3(2+r)}{N_c^2} \right] , \nonumber \\
Q_C \, (\Xi^{*0}) & = & +\frac{2r}{3} \left\{ B^\prime \left[ 1 -
\frac{3r(1+r)}{N_c} + \frac{3(1+r+r^2)}{N_c^2} \right]
\right. \nonumber \\ & & \hspace{2em} \left. - \frac{C^\prime}{N_c}
\left[ 1+r -\frac{3(1+r+2r^2)}{N_c} +\frac{6(1+r+r^2)}{N_c^2} \right]
\right\} , \\
Q_0 \, (\Xi^{*-}) & = & -B^\prime \, \frac{2(N_c^2+6N_c-9)}{3N_c^2} +
C^\prime \, \frac{4(N_c^2-3N_c-9)}{3N_c^3} , \nonumber \\
Q_Q \, (\Xi^{*-}) & = & -\frac{2r}{3} \left\{ B^\prime \left[ 1 +
\frac{3(1+r)}{N_c} - \frac{3(2+r)}{N_c^2} \right] -
\frac{2C^\prime}{N_c} \left[ 1 - \frac{3}{N_c} -\frac{3(r+2)}{N_c^2}
\right] \right\} , \nonumber \\
Q_C \, (\Xi^{*-}) & = &  -\frac{2r}{3} \left\{ B^\prime \left[ 1 +
\frac{3r(1+r)}{N_c} - \frac{3(1+r+r^2)}{N_c^2} \right]
\right. \nonumber \\ & & \left. - \frac{C^\prime}{N_c} \left[ (1+r) -
\frac{3(1+r)}{N_c} -\frac{6(1+r+r^2)}{N_c^2} \right] \right\} , \\
Q_0 \, (\Omega^-) & = & -B^\prime \, \frac{6(N_c-1)}{N_c^2} -
\frac{12C^\prime}{N_c^3} , \nonumber \\
Q_Q \, (\Omega^-) & = & -\frac{6r^2}{N_c} \left[ B^\prime \left( 1 -
\frac{1}{N_c} \right) + \frac{2C^\prime}{N_c^2} \right] , \nonumber
\\
Q_C \, (\Omega^-) & = & -\frac{6r^3}{N_c} \left[ B^\prime \left( 1 -
\frac{1}{N_c} \right) + \frac{2C^\prime}{N_c^2} \right] , \\
Q_{0, \, Q, \, C} \, (\Delta^+ p) & = & + \left( B^\prime -
\frac{2C^\prime}{N_c} \right) \sqrt{\frac{(N_c-1)(N_c+5)}{2N_c^2}} ,
\\
Q_{0, \, Q, \, C} \, (\Delta^0 n) & = & + \left( B^\prime -
\frac{2C^\prime}{N_c} \right) \sqrt{\frac{(N_c-1)(N_c+5)}{2N_c^2}} ,
\\
Q_0 \, (\Sigma^{*+} \Sigma^+) & = & + \left( B^\prime -
\frac{2C^\prime}{N_c} \right) \frac{N_c+5}{2N_c\sqrt{2}} , \nonumber
\\
Q_Q \, (\Sigma^{*+} \Sigma^+) & = & +\frac{B^\prime}{2\sqrt{2}} \left[
2-r + \! \frac{2+3r}{N_c} + \! \frac{8(1-r)}{N_c^2} \right] - \!
\frac{C^\prime}{N_c\sqrt{2}} \left[ 2-r - \! \frac{6-11r}{N_c}
+ \! \frac{8(1-r)}{N_c^2} \right] , \nonumber \\
Q_C \, (\Sigma^{*+} \Sigma^+) & = & +\frac{1}{2\sqrt{2}} \left\{
B^\prime \left[ 2-r + \frac{2-r+4r^2}{N_c} +
\frac{4(1-r)(2+r)}{N_c^2} \right] \right. \nonumber \\ & &
\hspace{3em} \left. - \frac{C^\prime}{N_c} \left[ 4-r-r^2 -
\frac{12-11r-11r^2}{N_c} + \frac{8(1-r)(2+r)}{N_c^2} \right] \right\}
, \\
Q_0 \, (\Sigma^{*0} \Sigma^0) & = & + \left( B^\prime -
\frac{2C^\prime}{N_c} \right) \frac{\sqrt{2}}{N_c} , \nonumber \\
Q_Q \, (\Sigma^{*0} \Sigma^0) & = & + \left( B^\prime -
\frac{2C^\prime}{N_c} \right) \frac{\sqrt{2}}{N_c} \left[ r +
\frac{2(1-r)}{N_c} \right] , \nonumber \\
Q_C \, (\Sigma^{*0} \Sigma^0) & = & +\frac{\sqrt{2}}{N_c} \left\{
B^\prime \left[ r^2 \! + \frac{(1-r)(2+r)}{N_c} \right] - \!
\frac{C^\prime}{N_c} \left[ r(1+r) + \frac{2(1-r)(2+r)}{N_c} \right]
\right\} , \\
Q_0 \, (\Sigma^{*0} \Lambda) & = & + \left( B^\prime -
\frac{2C^\prime}{N_c} \right) \sqrt{\frac{(N_c-1)(N_c+3)}{2N_c^2}} ,
\nonumber \\
Q_Q \, (\Sigma^{*0} \Lambda) & = & + r \left( B^\prime -
\frac{2C^\prime}{N_c} \right) \sqrt{\frac{(N_c-1)(N_c+3)}{2N_c^2}} ,
\nonumber \\
Q_C \, (\Sigma^{*0} \Lambda) & = & + r \left[ B^\prime -
\frac{C^\prime(1+r)}{N_c} \right] \sqrt{\frac{(N_c-1)(N_c+3)}{2N_c^2}}
, \\
Q_0 \, (\Sigma^{*-} \Sigma^-) & = & - \left( B^\prime -
\frac{2C^\prime}{N_c} \right) \frac{N_c-3}{2N_c\sqrt{2}} , \nonumber
\\
Q_Q \, (\Sigma^{*-} \Sigma^-) & = & -\frac{1}{2\sqrt{2}} \left\{
B^\prime \left[ 2-r + \frac{2-5r}{N_c} -\frac{8(1-r)}{N_c^2} \right]
\right. \nonumber \\ & & \hspace{3em} \left. - \frac{2C^\prime}{N_c}
\left[ 2-r -\frac{3(2-r)}{N_c} - \frac{8(1-r)}{N_c^2} \right] \right\}
, \nonumber \\
Q_C \, (\Sigma^{*-} \Sigma^-) & = & -\frac{1}{2\sqrt{2}} \left\{
B^\prime \left[ 2-r + \frac{2-r-4r^2}{N_c} -\frac{4(1-r)(2+r)}{N_c^2}
\right] \right. \nonumber \\ & & \hspace{3em} \left. -
\frac{C^\prime}{N_c} \left[ 4-r-r^2 -\frac{3(4-r-r^2)}{N_c} -
\frac{8(2+r)(1-r)}{N_c^2} \right] \right\} , \\
Q_0 \, (\Xi^{*0} \Xi^0) & = & + \left( B^\prime -
\frac{2C^\prime}{N_c} \right) \frac{\sqrt{2}(N_c+3)}{3N_c} , \nonumber
\\
Q_Q \, (\Xi^{*0} \Xi^0) & = & + \left( B^\prime -
\frac{2C^\prime}{N_c} \right) \frac{\sqrt{2}r}{3} \left[ 1 -
\frac{3(1-2r)}{N_c} + \frac{6(1-r)}{N_c^2} \right] , \nonumber \\
Q_C \, (\Xi^{*0} \Xi^0) & = & +\frac{\sqrt{2}r}{3} \left\{ B^\prime
\left[ 1 - \frac{3r(1-2r)}{N_c} + \frac{3(1-r)(1+2r)}{N_c^2} \right]
\right. \nonumber \\ & & \hspace{3em} \left. - \frac{C^\prime}{N_c}
\left[ 1+r -\frac{3(1+r-4r^2)}{N_c} +\frac{6(1-r)(1+2r)}{N_c^2}
\right] \right\} , \\
Q_0 \, (\Xi^{*-} \Xi^-) & = & - \left( B^\prime -
\frac{2C^\prime}{N_c} \right) \frac{\sqrt{2}(N_c-3)}{3N_c} ,
\nonumber \\
Q_Q \, (\Xi^{*-} \Xi^-) & = & -\frac{\sqrt{2}r}{3} \left\{ B^\prime
\left[ 1 + \frac{3(1-2r)}{N_c} - \frac{6(1-r)}{N_c^2} \right] -
\frac{2C^\prime}{N_c} \left[ 1 - \frac{3}{N_c} - \frac{6(1-r)}{N_c^2}
\right] \right\} , \nonumber \\
Q_C \, (\Xi^{*-} \Xi^-) & = & -\frac{\sqrt{2}r}{3} \left\{ B^\prime
\left[ 1 + \frac{3r(1-2r)}{N_c} - \frac{3(1-r)(1+2r)}{N_c^2} \right]
\right. \nonumber \\ & & \hspace{3em} \left. - \frac{C^\prime}{N_c}
\left[ 1+r - \frac{3(1+r)}{N_c} - \frac{6(1-r)(1+2r)}{N_c^2}
\right] \right\} .
\end{eqnarray}

\begin{table}
\caption{Values of $N_{u,d,s}$ [whence $\langle
\xbf{S}_\alpha^2 \rangle = \langle \alpha \alpha \rangle^{(0)} =
(N_\alpha/2)(N_\alpha/2 + 1)$] and matrix elements of the rank-0
tensors $\langle \alpha \beta \rangle^{(0)}$ with $\alpha \neq \beta$.
Since spin is unchanged by these operators, the matrix elements vanish
for all off-diagonal transitions except $\Sigma^0 \Lambda$; in that
case, the only nonvanishing entries are $\langle u s \rangle^{(0)} =
-\langle d s \rangle^{(0)} =
-\frac{1}{8}\sqrt{(N_c-1)(N_c+3)}$.\label{ab0}}
\begin{tabular}{c|c|c|c|c|c|c}
\hline\hline
State & $N_u$ & $N_d$ & $N_s$ & $\langle u d \rangle^{(0)}$ & $\langle
u s \rangle^{(0)}$ & $\langle d s \rangle^{(0)}$ \\
\hline\hline
$\Delta^{++}$ & $\frac{1}{2}(N_c+3)$ & $\frac{1}{2}(N_c-3)$ & 0 &
$-\frac{1}{16} (N_c-3)(N_c+7)$ & 0 & 0 \\
$\Delta^+$ & $\frac{1}{2}(N_c+1)$ & $\frac{1}{2}(N_c-1)$ & 0 &
$-\frac{1}{16} (N_c^2 + 4N_c - 29)$ & 0 & 0 \\
$\Delta^0$ & $\frac{1}{2}(N_c-1)$ & $\frac{1}{2}(N_c+1)$ & 0 &
$-\frac{1}{16} (N_c^2 + 4N_c - 29)$ & 0 & 0 \\
$\Delta^-$ & $\frac{1}{2}(N_c-3)$ & $\frac{1}{2}(N_c+3)$ & 0 &
$-\frac{1}{16} (N_c-3)(N_c+7)$ & 0 & 0 \\
$\Sigma^{*+}$ & $\frac{1}{2}(N_c+1)$ & $\frac{1}{2}(N_c-3)$ & 1 &
$-\frac{1}{16} (N_c-3)(N_c+5)$ & $+\frac{1}{16}(N_c+5)$ &
$-\frac{1}{16}(N_c-3)$ \\
$\Sigma^{*0}$ & $\frac{1}{2}(N_c-1)$ & $\frac{1}{2}(N_c-1)$ & 1 &
$-\frac{1}{16} (N_c^2 + 2N_c - 19)$ & $+\frac 1 4 $ & $+\frac 1 4$ \\
$\Sigma^{*-}$ & $\frac{1}{2}(N_c-3)$ & $\frac{1}{2}(N_c+1)$ & 1 &
$-\frac{1}{16} (N_c-3)(N_c+5)$ & $-\frac{1}{16}(N_c-3)$ &
$+\frac{1}{16}(N_c+5)$ \\
$\Xi^{*0}$ & $\frac{1}{2}(N_c-1)$ & $\frac{1}{2}(N_c-3)$ & 2 &
$-\frac{1}{16} (N_c-3)(N_c+3)$ & $+\frac{1}{12}(N_c+3)$ &
$-\frac{1}{12}(N_c-3)$ \\
$\Xi^{*-}$ & $\frac{1}{2}(N_c-3)$ & $\frac{1}{2}(N_c-1)$ & 2 &
$-\frac{1}{16} (N_c-3)(N_c+3)$ & $-\frac{1}{12}(N_c-3)$ &
$+\frac{1}{12}(N_c+3)$ \\
$\Omega^-$ & $\frac{1}{2}(N_c-3)$ & $\frac{1}{2}(N_c-3)$ & 3 &
$-\frac{1}{16} (N_c-3)(N_c+1)$ & 0 & 0 \\
$p$ & $\frac{1}{2}(N_c+1)$ & $\frac{1}{2}(N_c-1)$ & 0 & $-\frac{1}{16}
(N_c-1)(N_c+5)$ & 0 & 0 \\
$n$ & $\frac{1}{2}(N_c-1)$ & $\frac{1}{2}(N_c+1)$ & 0 & $-\frac{1}{16}
(N_c-1)(N_c+5)$ & 0 & 0 \\
$\Sigma^+$ & $\frac{1}{2}(N_c+1)$ & $\frac{1}{2}(N_c-3)$ & 1 &
$-\frac{1}{16} (N_c-3)(N_c+5)$ & $-\frac{1}{8}(N_c+5)$ &
$+\frac{1}{8}(N_c-3)$ \\
$\Sigma^0$ & $\frac{1}{2}(N_c-1)$ & $\frac{1}{2}(N_c-1)$ & 1 &
$-\frac{1}{16} (N_c^2 + 2N_c - 19)$ & $-\frac 1 2$ & $-\frac 1 2$ \\
$\Lambda$ & $\frac{1}{2}(N_c-1)$ & $\frac{1}{2}(N_c-1)$ & 1 &
$-\frac{1}{16} (N_c-1)(N_c+3)$ & 0 & 0 \\
%
%
$\Sigma^-$ & $\frac{1}{2}(N_c-3)$ & $\frac{1}{2}(N_c+1)$ & 1 &
$-\frac{1}{16} (N_c-3)(N_c+5)$ & $+\frac{1}{8} (N_c-3)$ &
$-\frac{1}{8} (N_c+5)$ \\
$\Xi^0$ & $\frac{1}{2}(N_c-1)$ & $\frac{1}{2}(N_c-3)$ & 2 &
$-\frac{1}{16} (N_c-3)(N_c+3)$ & $-\frac{1}{6} (N_c+3)$ &
$+\frac{1}{6} (N_c-3)$ \\
$\Xi^-$ & $\frac{1}{2}(N_c-3)$ & $\frac{1}{2}(N_c-1)$ & 2 &
$-\frac{1}{16} (N_c-3)(N_c+3)$ & $+\frac{1}{6} (N_c-3)$ &
$-\frac{1}{6} (N_c+3)$ \\
\hline
\end{tabular}
\end{table}

\begin{table}
\caption{Matrix elements of the operators $S_u^z$, $S_d^z$, and
$S_s^z$ in the state of maximal $S_z$.\label{S1}}
\begin{tabular}{c|c|c|c}
\hline\hline
State & $\langle S_u^z \rangle$ & $\langle S_d^z \rangle$ & $\langle
S_s^z \rangle$ \\
\hline\hline
$\Delta^{++}$ & $+\frac{3}{20}(N_c+7)$ & $-\frac{3}{20}(N_c-3)$ & 0 \\
$\Delta^+$ & $+\frac{1}{20}(N_c+17)$ & $-\frac{1}{20}(N_c-13)$ & 0 \\
$\Delta^0$ & $-\frac{1}{20}(N_c-13)$ & $+\frac{1}{20}(N_c+17)$ & 0 \\
$\Delta^-$ & $-\frac{3}{20}(N_c-3)$ & $+\frac{3}{20}(N_c+7)$ & 0 \\
$\Sigma^{*+}$ & $+\frac{1}{8}(N_c+5)$ & $-\frac{1}{8}(N_c-3)$ &
$+\frac 1 2$ \\
$\Sigma^{*0}$ & $+\frac 1 2$ & $+\frac 1 2$ & $+\frac 1 2$ \\
$\Sigma^{*-}$ & $-\frac{1}{8}(N_c-3)$ & $+\frac{1}{8}(N_c+5)$ &
$+\frac 1 2$ \\
$\Xi^{*0}$ & $+\frac{1}{12}(N_c+3)$ & $-\frac{1}{12}(N_c-3)$ & $+1$ \\
$\Xi^{*-}$ & $-\frac{1}{12}(N_c-3)$ & $+\frac{1}{12}(N_c+3)$ & $+1$ \\
$\Omega^-$ & 0 & 0 & $+\frac 3 2$ \\
$p$ & $+\frac{1}{12}(N_c+5)$ & $-\frac{1}{12}(N_c-1)$ & 0 \\
$n$ & $-\frac{1}{12}(N_c-1)$ & $+\frac{1}{12}(N_c+5)$ & 0 \\
$\Sigma^+$ & $+\frac{1}{12}(N_c+5)$ & $-\frac{1}{12}(N_c-3)$ & $-\frac
1 6$ \\
$\Sigma^0$ & $+\frac 1 3$ & $+\frac 1 3$ & $-\frac 1 6$ \\
$\Lambda$ & 0 & 0 & $+\frac 1 2$ \\
$\Sigma^-$ & $-\frac{1}{12}(N_c-3)$ & $+\frac{1}{12}(N_c+5)$ & $-\frac
1 6$ \\
$\Xi^0$ & $-\frac{1}{36}(N_c+3)$ & $+\frac{1}{36}(N_c-3)$ & $+\frac 2
3$ \\
$\Xi^-$ & $+\frac{1}{36}(N_c-3)$ & $-\frac{1}{36}(N_c+3)$ & $+\frac 2
3$ \\
$\Delta^+ p$ & $+\frac{1}{6\sqrt{2}} \sqrt{(N_c-1)(N_c+5)}$ &
$-\frac{1}{6\sqrt{2}} \sqrt{(N_c-1)(N_c+5)}$ & 0 \\
$\Delta^0 n$ & $+\frac{1}{6\sqrt{2}} \sqrt{(N_c-1)(N_c+5)}$ &
$-\frac{1}{6\sqrt{2}} \sqrt{(N_c-1)(N_c+5)}$ & 0 \\
$\Sigma^{*+} \Sigma^+$ & $+\frac{1}{12\sqrt{2}} (N_c+5)$ &
$-\frac{1}{12\sqrt{2}} (N_c-3)$ & $-\frac{\sqrt{2}}{3}$ \\
$\Sigma^{*0} \Sigma^0$ & $+\frac{1}{3\sqrt{2}}$ &
$+\frac{1}{3\sqrt{2}}$ & $-\frac{\sqrt{2}}{3}$ \\
$\Sigma^{*0} \Lambda$ & $+\frac{1}{6\sqrt{2}} \sqrt{(N_c-1)(N_c+3)}$ &
$-\frac{1}{6\sqrt{2}} \sqrt{(N_c-1)(N_c+3)}$ & 0 \\
$\Sigma^{*-} \Sigma^-$ & $-\frac{1}{12\sqrt{2}} (N_c-3)$ &
$+\frac{1}{12\sqrt{2}} (N_c+5)$ & $-\frac{\sqrt{2}}{3}$ \\
$\Xi^{*0} \Xi^0$ & $+\frac{1}{9\sqrt{2}} (N_c+3)$ &
$-\frac{1}{9\sqrt{2}} (N_c-3)$ & $-\frac{\sqrt{2}}{3}$ \\
$\Xi^{*-} \Xi^-$ & $-\frac{1}{9\sqrt{2}} (N_c-3)$ &
$+\frac{1}{9\sqrt{2}} (N_c+3)$ & $-\frac{\sqrt{2}}{3}$ \\
$\Sigma^0 \Lambda$ & $-\frac{1}{4\sqrt{6}} \sqrt{(N_c-1)(N_c+3)}$ &
$+\frac{1}{4\sqrt{6}} \sqrt{(N_c-1)(N_c+3)}$ & 0 \\
\hline
\end{tabular}
\end{table}

\begin{table}
\caption{The matrix elements $\langle \alpha \alpha \rangle^{(2)}$ for
all relevant states.\label{aa2}}
\begin{tabular}{c|c|c|c}
\hline\hline
State & $\langle u u \rangle^{(2)}$ & $\langle d d \rangle^{(2)}$ &
$\langle s s \rangle^{(2)}$ \\
\hline\hline
$\Delta^{++}$ & $+\frac{1}{40}(N_c+7)(N_c+9)$ &
$+\frac{1}{40}(N_c-5)(N_c-3)$ & 0 \\
$\Delta^+$ & $-\frac{1}{40}(N_c-7)(N_c+7)$ &
$-\frac{1}{40}(N_c-3)(N_c+11)$ & 0 \\
$\Delta^0$ & $-\frac{1}{40}(N_c-3)(N_c+11)$ &
$-\frac{1}{40}(N_c-7)(N_c+7)$ & 0 \\
$\Delta^-$ & $+\frac{1}{40}(N_c-5)(N_c-3)$ &
$+\frac{1}{40}(N_c+7)(N_c+9)$ & 0 \\
$\Sigma^{*+}$ & $+\frac{1}{80}(N_c+5)(N_c+7)$ &
$+\frac{1}{80}(N_c-5)(N_c-3)$ & 0 \\
$\Sigma^{*0}$ & $-\frac{1}{40}(N_c-3)(N_c+5)$ &
$-\frac{1}{40}(N_c-3)(N_c+5)$ & 0 \\
$\Sigma^{*-}$ & $+\frac{1}{80}(N_c-5)(N_c-3)$ &
$+\frac{1}{80}(N_c+5)(N_c+7)$ & 0 \\
$\Xi^{*0}$ & 0 & 0 & $+1$ \\
$\Xi^{*-}$ & 0 & 0 & $+1$ \\
$\Omega^-$ & 0 & 0 & $+3$ \\
$\Delta^+ p$ & $+\frac{1}{20\sqrt{2}}(N_c+7) \sqrt{(N_c-1)(N_c+5)}$ &
$+\frac{1}{20\sqrt{2}}(N_c-3) \sqrt{(N_c-1)(N_c+5)}$ & 0 \\
$\Delta^0 n$ & $-\frac{1}{20\sqrt{2}}(N_c-3) \sqrt{(N_c-1)(N_c+5)}$ &
$-\frac{1}{20\sqrt{2}}(N_c+7) \sqrt{(N_c-1)(N_c+5)}$ & 0 \\
$\Sigma^{*+} \Sigma^+$ & $+\frac{1}{40\sqrt{2}} (N_c+5)(N_c+7)$ &
$+\frac{1}{40\sqrt{2}}(N_c-5)(N_c-3)$ & 0 \\
$\Sigma^{*0} \Sigma^0$ & $-\frac{1}{20\sqrt{2}} (N_c-3)(N_c+5)$ &
$-\frac{1}{20\sqrt{2}} (N_c-3)(N_c+5)$ & 0 \\
$\Sigma^{*0} \Lambda$ & 0 & 0 & 0 \\
$\Sigma^{*-} \Sigma^-$ & $+\frac{1}{40\sqrt{2}} (N_c-5)(N_c-3)$ &
$+\frac{1}{40\sqrt{2}} (N_c+5)(N_c+7)$ & 0 \\
$\Xi^{*0} \Xi^0$ & 0 & 0 & $-\sqrt{2}$ \\
$\Xi^{*-} \Xi^-$ & 0 & 0 & $-\sqrt{2}$ \\
\hline
\end{tabular}
\end{table}

\begin{table}
\caption{The matrix elements $\langle \alpha \beta \rangle^{(2)}$, with
$\alpha \neq \beta$, for all relevant states.\label{ab2}}
\begin{tabular}{c|c|c|c}
\hline\hline
State & $\langle u d \rangle^{(2)}$ & $\langle u s \rangle^{(2)}$ &
$\langle d s \rangle^{(2)}$ \\
\hline\hline
$\Delta^{++}$ & $-\frac{1}{40} (N_c-3)(N_c+7)$ & 0 & 0\\
$\Delta^+$ & $+\frac{1}{40} (N_c^2+4N_c+19)$ & 0 & 0 \\
$\Delta^0$ & $+\frac{1}{40} (N_c^2+4N_c+19)$ & 0 & 0 \\
$\Delta^-$ & $-\frac{1}{40} (N_c-3)(N_c+7)$ & 0 & 0 \\
$\Sigma^{*+}$ & $-\frac{1}{80} (N_c-3)(N_c+5)$ & $+\frac{1}{8}(N_c+5)$
& $-\frac{1}{8}(N_c-3)$ \\
$\Sigma^{*0}$ & $+\frac{1}{40} (N_c^2+2N_c+5)$ & $+\frac 1 2$ &
$+\frac 1 2$ \\
$\Sigma^{*-}$ & $-\frac{1}{80} (N_c-3)(N_c+5)$ & $-\frac{1}{8}
(N_c-3)$ & $+\frac{1}{8} (N_c+5)$ \\
$\Xi^{*0}$ & 0 & $+\frac{1}{6} (N_c+3)$ & $-\frac{1}{6} (N_c-3)$ \\
$\Xi^{*-}$ & 0 & $-\frac{1}{6} (N_c-3)$ & $+\frac{1}{6} (N_c+3)$ \\
$\Omega^-$ & 0 & 0 & 0 \\
$\Delta^+ p$ & $-\frac{1}{20\sqrt{2}} (N_c+2) \sqrt{(N_c-1)(N_c+5)}$ &
0 & 0 \\
$\Delta^0 n$ & $+\frac{1}{20\sqrt{2}} (N_c+2) \sqrt{(N_c-1)(N_c+5)}$ &
0 & 0 \\
$\Sigma^{*+} \Sigma^+$ & $-\frac{1}{40\sqrt{2}} (N_c-3)(N_c+5)$ &
$-\frac{1}{8\sqrt{2}} (N_c+5)$ & $+\frac{1}{8\sqrt{2}} (N_c-3)$ \\
$\Sigma^{*0} \Sigma^0$ & $+\frac{1}{20\sqrt{2}} (N_c^2+2N_c+5)$ &
$-\frac{1}{2\sqrt{2}}$ & $-\frac{1}{2\sqrt{2}}$ \\
$\Sigma^{*0} \Lambda$ & 0 & $+\frac{1}{4\sqrt{2}}
\sqrt{(N_c-1)(N_c+3)}$ & $-\frac{1}{4\sqrt{2}} \sqrt{(N_c-1)(N_c+3)}$
\\
$\Sigma^{*-} \Sigma^-$ & $-\frac{1}{40\sqrt{2}} (N_c-3)(N_c+5)$ &
$+\frac{1}{8\sqrt{2}} (N_c-3)$ & $-\frac{1}{8\sqrt{2}} (N_c+5)$ \\
$\Xi^{*0} \Xi^0$ & 0 & $+\frac{1}{6\sqrt{2}} (N_c+3)$ &
$-\frac{1}{6\sqrt{2}} (N_c-3)$ \\
$\Xi^{*-} \Xi^-$ & 0 & $-\frac{1}{6\sqrt{2}} (N_c-3)$ &
$+\frac{1}{6\sqrt{2}} (N_c+3)$ \\
\hline
\end{tabular}
\end{table}

\end{document}